\documentclass{osa-article}

\journal{oe}

\articletype{OE}
\usepackage{graphicx}
\usepackage{epsfig}

\usepackage[]{graphicx}
\usepackage{engord}
\usepackage{multirow}
\usepackage{amsmath}
\usepackage{amssymb}
\usepackage{psfrag}
\usepackage{longtable}
\usepackage{bm}
\usepackage{array}
\usepackage{setspace}
\usepackage{exscale}
\usepackage{relsize}
\usepackage{amsmath}
\usepackage{wasysym}
\begin{document}

\title{Compressive spectral imaging based on hexagonal blue noise coded apertures}

\author{Hao Zhang,\authormark{1} Xu Ma,\authormark{1,*}, Daniel L. Lau\authormark{2}, and Gonzalo R. Arce\authormark{3}}

\address{\authormark{1}Key Laboratory of Photoelectronic Imaging Technology and System of Ministry of Education of China, School of Optics and Photonics, Beijing Institute of Technology, Beijing, 100081, China, P. R.\\
	 \authormark{2}Department of Electrical and Computer Engineering, University of Kentucky, Lexington, KY, 40506, USA \\
	\authormark{3}Department of Electrical and Computer Engineering, University of Delaware, Newark, DE, 19716, USA}

\email{\authormark{*}maxu@bit.edu.cn}




\begin{abstract}
The coded aperture snapshot spectral imager (CASSI) is a computational imaging system that acquires a three dimensional (3D) spectral data cube by a single or a few two dimensional (2D) measurements. The 3D data cube is reconstructed computationally. Binary on-off random coded apertures with square pixels are primarily implemented in CASSI systems to modulate the spectral images in the image plane. The design and optimization of coded apertures have been shown to improve the imaging performance of these systems significantly. This work proposes a different approach to code design. Instead of using traditional squared tiled coded elements, hexagonal tiled elements are used. The dislocation between the binary hexagonal coded apertures and the squared detector pixels is shown to introduce an equivalent grey-scale spatial modulation that increases the degrees of freedom in the sensing matrix, thus further improving the spectral imaging performance. Based on the restricted isometry property (RIP) of compressive sensing theory, this paper proves that optimal coded aperture patterns under a hexagonal lattice obey blue noise spatial characteristic, where ``on'' elements are placed as far from each other as possible. In addition, optimal coded apertures used in different snapshots are complementary to each other. This paper also proves the superiority of the hexagonal blue noise coded aperture over the traditional coded apertures with squared tiled elements. Based on a set of simulations, the proposed hexagonal tiled coded apertures are shown to effectively improve the imaging performance of CASSI systems compared to that of traditional coded apertures on rectangular lattices. 
\end{abstract}

\section{Introduction}
\label{sec1}
\begin{figure}[!h]
	\centering\includegraphics[width=13.5cm]{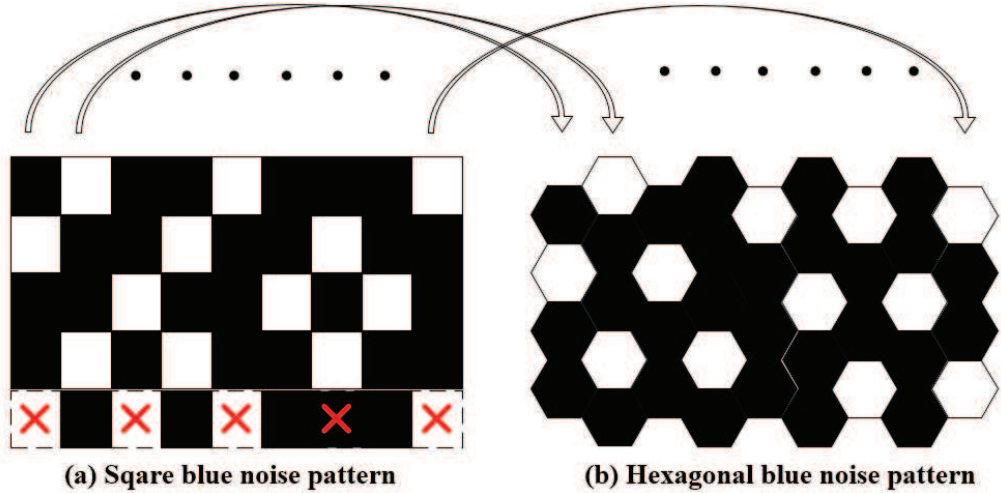} \caption{Sketch of the CASSI system with the proposed hexagonal blue noise coded aperture.}
	\label{figure1}
\end{figure}

Hyperspectral imaging systems acquire large amounts of data in a given spectral range to construct a three-dimensional (3D) spatiospectral data cube\cite{introduction}. Conventional spectral imagers include whisk broom scanners, push broom scanners and staring imagers, all of which require a time-consuming scanning process along either the spatial or spectral coordinates\cite{tunable}. With the improvement of the spatial and spectral resolutions, the data throughput of hyperspectral imagers dramatically increases, which poses great challenges on conventional spectral imagers in terms of data acquisition, storage and transmission. To circumvent these limitations, compressive spectral imaging methods have been recently developed to acquire high-dimensional hyperspectral data using a set of focal plane array (FPA) measurements\cite{introduction}. Wagadarikar et al. proposed the coded aperture snapshot spectral imager (CASSI), which captures the 3D spectral data cube with just a single two-dimensional (2D) measurement \cite{Spectral}. As shown in Fig.~\ref{figure1}, the spectral data cube of the target is first modulated by a block/unblock coded aperture in the image plane, and then different spectral slices are laterally shifted by a dispersive element before they are collapsed and multiplexed onto the FPA detector. Based on the strong correlation across the spectra, and the sparse representation of the spectral images on a pre-defined basis, the 3D spectral data cube can be reconstructed from compressive measurements using iterative optimization algorithms\cite{TwIST,GPSR}. 

Assume the hyperspectral data cube consists of $N \times M $ spatial pixels and $L$ spectral bands. When the resolution of the coded aperture is the same as the detector, the data cube can be reconstructed from just $N \times (M + L - 1)$ measurements by solving a regularized inverse problem \cite{Spectral}. The broadening in the x-axis measurements is induced by the dispersion element. In some cases, a single snapshot is not sufficient to capture enough spectral information to obtain an accurate reconstruction. In order to overcome this limitation, multiple snapshots are usually taken to ameliorate the condition of the underdetermined inverse problem, where each snapshot uses a different coded aperture \cite{Multiframe,agile,Development,RGB_image_sensors}. For $K$ snapshots, the total amount of compressive measurements increases to $K \times N \times (M + L - 1)$.

In addition to the number of measurements, the structure of the coded apertures has a significant influence on the reconstruction quality of spectral images. Initially, random binary coded apertures were used in CASSI. Random coded apertures are convenient to implement but suboptimal in the compressive sensing (CS) reconstruction framework since the sensing matrix in CASSI systems are highly structured and sparse\cite{agile,Development}. In order to further improve the sensing efficiency, a set of optimization approaches were proposed to design coded aperture patterns based on CS theory\cite{mismatched,Side,higher,Unmixing}. The restricted isometry property (RIP) provides a design metric for coded apertures as it guarantees the probability of accurate image reconstruction. Based on the RIP condition, these approaches optimized the coded apertures to obtain improved sensing matrices, which retained the incoherence between the projection matrix and the sparse basis\cite{colored,Restricted,classification,Rank}. In all of these CASSI systems, binary coded apertures with block or unblock elements in rectangular tilings are used so as to match the focal plane array tessellation. In order to improve modulation freedom and increase the dynamic range of the FPA sensor, grey-scale coded apertures have been proposed using digital-micromirror-devices (DMD), taking advantage of the fast switching time of the micro-mirrors, which enables the use of a pulse-width modulation techniques for the synthesis of grey-scale values\cite{High-dynamic}. Recently, colored coded apertures, which physically allow the pass of specific bandwidths of different spatial locations, have been introduced to improve the reconstruction quality\cite{Snapshot,colored}.

All of these aforementioned methods, however, are limited to squared tiled coded apertures (${\rm CA}_{\Square}$). It is thus natural to ask if different tiling geometries can provide improved coding strategies in CASSI. Indeed, hexagonal coded apertures (${\rm CA}_{\hexagon}$) are explored in this paper to provide several advantages in the modulation of spectral images. In the past, hexagonal lattices have been used in the processing of multidimensional bandlimited signals. The hexagonal lattice was shown to possess some advantages compared to rectangular lattices with respect to sampling density\cite{Hexagonal,comparison,hexagonal grids,display}. 
As shown in Fig.~\ref{figure1}, the proposed method grids the coded aperture into honeycomb-like arrays with hexagonal elements instead of square pixels. The proposed ${\rm CA}_{\hexagon}$ have the following two merits. First, the geometric dislocation between the binary hexagonal coded apertures and the square pixels on detector introduces an equivalent grey-scale spatial modulation on the spectral images, thus increasing the degrees of freedom in the sensing matrix. Moreover, the transmittance of each hexagonal element is assigned to be binary, and the blocked/unblocked coded aperture patterns can be easily implemented by binary lithography masks. Thus, the proposed binary ${\rm CA}_{\hexagon}$ are cost efficient compared to other grey-scale coded aperture implementations, which may require DMD temporary dithering, or grey-scale lithography. 

Another contribution of this paper is to derive the optimal distribution of the ${\rm CA}_{\hexagon}$. Based on the RIP criterion, this paper proves that optimal coded aperture patterns with hexagonal tiled elements obey blue noise distributions under a hexagonal lattice. The spatial and spectral characteristic of blue noise and other spectrally shape dithering patterns can be found in \cite{blue_noise_square,multitone dithering,Green-noise digital,Digital halftoning,Minimizing stochastic}. In addition, the optimal ${\rm CA}_{\hexagon}$ used in different snapshots should be complementary to each other. This paper also proves the superiority of the blue noise ${\rm CA}_{\hexagon}$ over the traditional random ${\rm CA}_{\Square}$ or blue noise ${\rm CA}_{\Square}$. A set of simulations are conducted to verify the improvement of imaging performance obtained by the proposed coding strategy. The influence of the offset between the ${\rm CA}_{\hexagon}$ and detector on the imaging performance is also studied.

The remainder of this paper is organized as follows. Modelling of the general CASSI system is described in Section~\ref{sec2}. Design methods and the theoretical proof of the blue noise ${\rm CA}_{\hexagon}$ are provided in Section~\ref{sec3}. Simulations and analysis are presented in Section~\ref{sec4}. Conclusions are provided in Section~\ref{sec5}.

\section{Modelling of the CASSI system}
\label{sec2}

\subsection{Forward imaging model of CASSI system }
\label{sec2.1}

 Let ${f_0}(x,y,\lambda)$ represent the spectral data of the target, where $x$, $y$ are the spatial coordinates and $\lambda $ is the spectral coordinate. The incident spectral images are first modulated in the spatial domain by the coded aperture whose transmission function is denoted by $T(x,y)$. Subsequently, the coded spectral image planes are shifted along the lateral direction by the prism, and then integrated along the $\lambda $ axis on the FPA detector. Thus, the measurements on the detector can be written as\cite{Rank}
\begin{equation}\label{e2-1}
{f_1}(x,y) = \mathlarger{\int}\mathlarger{\int}\mathlarger{\int} T(x,y){f_0}(x,y,\lambda )h(x' - \alpha \lambda  - x,y' - y)dx'dy'd\lambda,
\end{equation}
where $h(x' - \alpha \lambda  - x,y' - y)$ is the optical impulse response of the CASSI system. Assume the prism induces a linear dispersion effect, where $\alpha$ is the linear dispersion rate of the prism.

Due to the pixelated nature of the detector array, the continuous model in Eq.~(\ref{e2-1}) is first discretized. The discretized output at the detector corresponding to the $k\mathrm{th}$ coded  aperture ${\bf{T}}^k$ is given by
\begin{equation}\label{e2-2}
{\bf{Y}}_{ij}^k = \sum\limits_{l = 1}^L {{{\bf{F}}_{i,j + l,l}}} {\bf{T}}_{_{i,j + l}}^k + \omega _{ij}^k,
\end{equation}
where ${{\bf{Y}}_{ij}^k}$ is the measurement on the $(i,j)\mathrm{th}$ detector pixel at the $k\mathrm{th}$ snapshot\cite{Multiframe,agile}. The dimension of ${{\bf{Y}}^k}$ is $N \times (M+L-1)$. The dimension of ${\bf{T}}^k$ is $N \times M$, and ${\bf{T}}_{ij}^k$ is the $(i,j)\mathrm{th}$ pixel on the coded aperture. ${{\bf{F}}}$ is the 3D spectral data cube of the target with dimension $N \times M \times L$. ${{\bf{F}}_{ijl}}$ is the voxel in the data cube at the spatial coordinate $(i,j)$ and the $l\mathrm{th}$ spectral band. $\omega _{ij}^k$ is the measurement noise on the detector. The compressive measurement at the $k\mathrm{th}$ snapshot can be written in the following matrix notation:
\begin{equation}\label{e2-3}
{{\bm{y}}^k}={{\mathbf{H}}^k}{\bm{f + }}{{\bm{\omega }}^k},
\end{equation}
where $ {{\bm{y}}^k} \in {\mathbb{R}^{N(M + L - 1) \times 1}}$ is a vector concatenating all of the measurements $\mathbf{Y}_{ij}^k $ in Eq.~(\ref{e2-2}). $ {{\bf{H}}^k} $ is the system matrix representing the effect of the $k\mathrm{th}$ coded aperture and the dispersion effect realized by the prism. $\bm{f}$ is the vectorized representation of ${\bf{F}}$. ${\bm{\omega }}^k$ is the sensing noise in the CASSI system at the $k\mathrm{th}$ snapshot. Taking into account multiple snapshots, the measurements can be concatenated together, and the forward imaging model becomes
\begin{equation}\label{e2-4}
\bm{y} = \mathbf{H}\bm{f} + \bm{\omega},
\end{equation}
where ${{\bm{y}}} = [({\bm{y}^1})^T,({\bm{y}^2})^T,\ldots,({\bm{y}^K})^T]^T$ and $\mathbf{H}=[({\mathbf{H}^1})^T,({\mathbf{H}^2})^T,\ldots,({\mathbf{H}^K})^T]^T$\cite{Multiframe,agile,Development}. 
Suppose the data cube is highly correlated across the spatial and spectral domains, and is sparse in some representation basis $\mathbf{\Psi}$.
Then, $\bm{f}$ in Eq.~(\ref{e2-4}) can be represented as
\begin{equation}\label{e2-5}
\bm{f} =\mathbf{\Psi} \bm{\theta},
\end{equation}
 where $\mathbf{\Psi}  = ({\mathbf{\Psi} _1} \otimes {\mathbf{\Psi} _2}) \in {R^{NML \times NML}}$ is a 3D representation basis for the data cube, $\mathbf{\Psi} _1$ is the 2D wavelet Symmlet-8 basis to depict the correlation in spatial domain, $\mathbf{\Psi} _2$ is the one-dimensional (1D) DCT basis in spectral domain, $\otimes$ is the Kronecker product, and $\bm{\theta}$ is the coefficient vector in the 3D basis. Substituting Eq.~(\ref{e2-5}) into Eq.~(\ref{e2-4}), we have
\begin{equation}\label{e2-6}
\bm{y} = \mathbf{H}\mathbf{\Psi} \bm{\theta} + \bm{\omega}.
\end{equation}

\begin{figure}[!h]
	\centering\includegraphics[width=10cm]{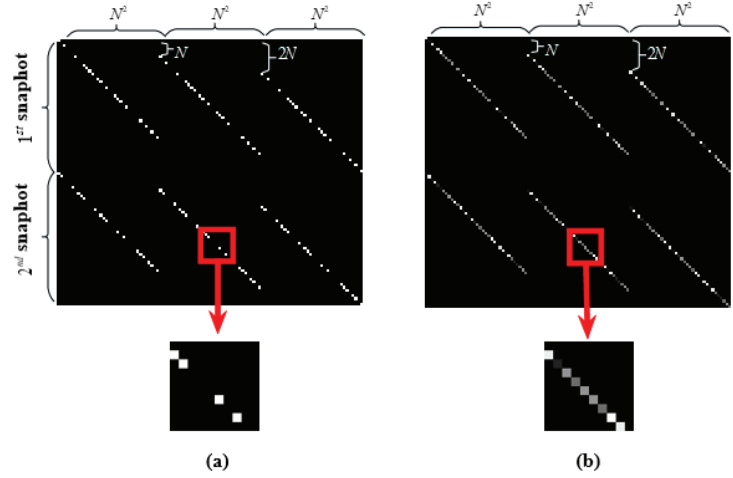} \caption{Structure of the $\mathbf{H}$ matrices ($K = 2$, $N=M = 6$, $L = 3$) for (a) the random ${\rm CA}_{\Square}$ and (b) the blue noise ${\rm CA}_{\hexagon}$. Notice that entries in (a) are either 0 or 1, while the entry values in (b) vary in the interval [0, 1], and provide more degrees of freedom on the number of elements in  $\mathbf{H}$ that are being coded, and on the quantization levels of the coder.}
	\label{figure2}
\end{figure}

Notice that ${\bf{A}} = {\bf{H\Psi }} \in {R^{Q_1 \times Q_2}}$ is the sensing matrix, where $Q_1=KV$, $V=N(M+L-1)$, and $Q_2=NML$. 
In the conventional CASSI, the sensing matrix $\bf{A}$ is determined by the matrix $\bf{H}$, which has the structure shown in Fig.~\ref{figure2}(a). In this example, $K=2$, $N=M=6$, $L=3$. The coded apertures have binary random patterns with square pixels, which are referred to as random ${\rm CA}_{\Square}$. The entries of the coded apertures obey the Bernoulli distribution, and the transmittance is 50\%. It can be observed that the $\mathbf{H}$ matrix is sparse and highly structured, which consists of a set of diagonal patterns determined by the coded aperture entries ${\bf{T}}_{ij}^k$, and repeated in the horizontal direction\cite{introduction}. The sensing matrix $\mathbf{A}$ plays a crucial role in the mathematics of the inverse CS problem. Thus, our main task is to design the coded apertures ${\bf{T}}^k$, such that the sensing matrix $\bf{A}$ is better conditioned based on the RIP to improve the reconstruction quality.


\subsection{Reconstruction model of spectral images}
\label{sec2.2}
According to CS theory, the sparse signal can be recovered from fewer samples than those required by the Shannon-Nyquist sampling theorem \cite{E.Candes,D.Donoho,sampling}.
In the CASSI system, the compressive measurements on the FPA are used to reconstruct the spectral images of the target. According to the imaging model in Eq.~(\ref{e2-6}), the sparse coefficients of the spectral data cube can be reconstructed by solving the following $l_1$-norm minimization problem: 
\begin{equation}\label{e3-6}
\hat{\bm{\theta}} = \arg\min \limits_{\bm{\theta}} \|\bm{\theta}\|_1,~s.t.~\|\bm{y} - \mathbf{H}  \mathbf{\Psi} \bm{\theta} \|^2_2 < \epsilon,
\end{equation}
where $\epsilon$ is a small positive parameter used to constrain the upper bound of the reconstruction error, and $\|\cdot\|_1$ and $\|\cdot\|_2$ represent the $l_1$-norm and $l_2$-norm, respectively. In this paper, the gradient projection for sparse reconstruction
(GPSR) algorithm is used to solve for the above optimization problem \cite{GPSR}. Other algorithms developed in the CS realm can also be used \cite{Rice}. Finally, the spectral images can be recovered as $\hat{\bm{f}} = \bm{\Psi} \hat{\bm{\theta}}$, where $\hat{\bm{\theta}}$ is the solution of Eq.~(\ref{e3-6}).

\section{Hexagonal blue noise coding strategy}
\label{sec3}
This section introduces hexagonal blue noise coding strategies in details. Section~\ref{sec3.1} describes the design method of the ${\rm CA}_{\hexagon}$, and its equivalent grey-scale coded aperture. Section~\ref{sec3.2} proves the optimality of the ${\rm CA}_{\hexagon}$ with complementary blue noise sampling patterns. Section~\ref{sec3.3} will provide a method to generate blue noise ${\rm CA}_{\hexagon}$.
\subsection{Design method of ${\rm CA}_{\hexagon}$}
\label{sec3.1}
\begin{figure}[!h]
	\centering\includegraphics[width=7cm]{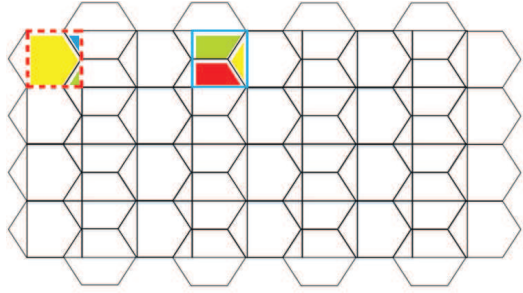} \caption{The periodic correspondence between the square pixel array and ${\rm CA}_{\hexagon}$. The red dotted block and blue solid block represent the two different mapping relationships between the square pixels and ${\rm CA}_{\hexagon}$.  }
	\label{figure4}
\end{figure}

Mersereau showed that for a 2D signal with circular or isotropic spectral support, the hexagonal sampling lattice is optimal. It requires fewer samples than the minimal rectangularly sampled signal\cite{Hexagonal,comparison}.
As shown in Fig.~\ref{figure4}, the proposed ${\rm CA}_{\hexagon}$ is gridded into a honeycomb-like array with hexagonal elements instead of square pixels. On the other hand, pixels on the target and detector are square. Due to the different sampling lattices used by the coded apertures and the detector array, a geometric dislocation between the coded aperture elements and the pixels on the target and detector exists. The binary ${\rm CA}_{\hexagon}$ turns out to be mathematically equivalent to grey-scale ${\rm CA}_{\Square}$ as will be shown shortly. The transmittance of each equivalent grey-scale pixel is the weighted average of several adjacent elements on the binary ${\rm CA}_{\hexagon}$. This increases the degrees of freedom in the spatial modulation, thus improving the quality of the reconstructed images. From a practical point of view, it is cost efficient to emulate the grey-scale coded apertures using the binary lithography mask with hexagonal tiled elements. Furthermore, state-of-the-art DMD devices can realize diamond-shaped micro-mirrors aligned in a hexagonal lattice\cite{bostonmicromachines}. This kind of DMD can be also used to implement the proposed ${\rm CA}_{\hexagon}$, and different coded aperture patterns can be easily switched by changing the reflective angles of the micro-mirrors.

Although the geometric dislocation exists between the ${\rm CA}_{\hexagon}$ and square detector array, there is a periodic correspondence between them. Consider first a special case where the left boundary of the square detector array is aligned with the most left vertices of the ${\rm CA}_{\hexagon}$, as shown in Fig.~\ref{figure4}. It is observed that each square pixel on the spectral images will be modulated by three adjacent hexagonal elements. The transmittance within one square pixel region thus equals the weighted average of the transmittance coefficients corresponding to the three adjacent binary hexagonal elements. In the rectangular coordinate system, the binary ${\rm CA}_{\hexagon}$ are thus mathematically equivalent to grey-scale coded apertures with squared tiled elements.  

As shown in Fig.~\ref{figure4}, there are only two types of mapping relationship between one square pixel and its adjacent hexagonal elements, which happen in the odd column and even column, respectively. Hereafter, we refer these two types of relationship as Type I and Type II. In Fig.~\ref{figure4}, the pixels surrounded by red dotted line and blue solid line show the examples of Type I and Type II, respectively. For Type I, one pixel on the square array is contributed by three adjacent hexagonal elements, the first one is on the left (yellow), the second one is in the top right corner (blue), and the third one is in the bottom right corner (green). The percentages of the hexagonal elements spanned by the yellow area (left), blue area (top right corner) and green area (bottom right corner) are $1-\sqrt{3}/12$, $\sqrt{3}/24$ and $\sqrt{3}/24$, respectively. Let $\mathbf{T}_b$ and $\mathbf{T}_g$ respectively indicate the transmittance of the binary hexagonal coded aperture and that of the equivalent grey-scale square coded aperture. Then, the transmittance of the $(i,j)$th pixel of $\mathbf{T}_g$ can be calculated as 
\begin{equation}\label{e3-1}
\mathbf{T}_g(i,j) = (1 - \frac{\sqrt 3}{12})\mathbf{T}_b(i,j) + \frac{\sqrt 3}{24}\mathbf{T}_b(i,j + 1) + \frac{\sqrt 3}{24}\mathbf{T}_b(i + 1,j + 1),
\end{equation}
where $\mathbf{T}_b(i,j)$, $\mathbf{T}_b(i,j + 1)$ and $\mathbf{T}_b(i + 1,j + 1)$ represent the $(i,j)$th, $(i,j+1)$th and $(i+1,j+1)$th elements in $\mathbf{T}_b$. 

Similarly, for Type II, one pixel of the square array is also modulated by three adjacent hexagonal elements, the first one is in the top left corner (green), the second one is in the bottom left corner (red), and the third one is on the right (yellow). The percentages of the hexagonal elements spanned by the green area (top left corner), red area (bottom left corner) and  yellow area (right) are $1/2 - \sqrt 3 /24$, $1/2 - \sqrt 3 /24$ and $\sqrt 3 / 12$, respectively. Then, the transmittance of the $(i,j)$th pixel of $\mathbf{T}_g$ can be calculated as  
\begin{equation}\label{e3-2}
\mathbf{T}_g(i,j) = (\frac{1}{2}-\frac{\sqrt 3}{24})\mathbf{T}_b(i,j) + (\frac{1}{2} - \frac{\sqrt 3 }{24})\mathbf{T}_b(i+1,j) + \frac{\sqrt 3}{12}\mathbf{T}_b(i,j + 1).
\end{equation}

\begin{figure}[!h]
	\centering\includegraphics[width=7cm]{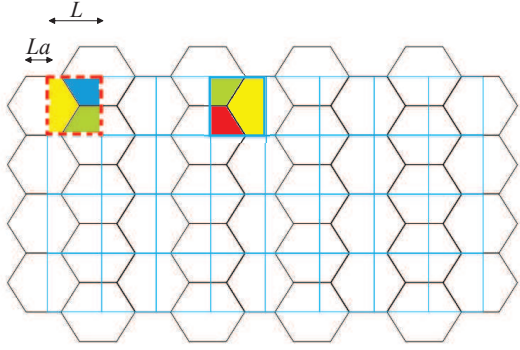} \caption{The periodic correspondence between the moved square detector array and the ${\rm CA}_{\hexagon}$.}
	\label{figure5}
\end{figure}

Next, consider a more general case, where the square detector array is moved along the $x$-axis with respect to the ${\rm CA}_{\hexagon}$. In this case, the transmittance of the equivalent grey-scale ${\rm CA}_{\square}$ changes according to the displacement of the square detector array. As shown in Fig.~\ref{figure5}, suppose the side length of the square pixel is $L$, and assume the offset between the square array and the ${\rm CA}_{\hexagon}$ is $aL$, where $a$ is the ratio of offset to the pixel side length. The overlapping modes between the square array and ${\rm CA}_{\hexagon}$ can be still classified into two types, as shown by the red dotted block and blue solid block in Fig.~\ref{figure5}, respectively.
For Type I, the percentages of the hexagonal elements spanned by the yellow area (left), blue area (top right corner) and green area (bottom right corner) are $1 -\sqrt 3 / 12 - a$, $\sqrt 3 / 24 + a/2$ and $\sqrt 3 / 24 + a/2$, respectively. Then, the transmittance of the $(i,j)$th pixel of $\mathbf{T}_g$ can be calculated as  
\begin{equation}\label{e3-3}
\mathbf{T}_g(i,j) = (1 - \frac{\sqrt 3}{12} - a)\mathbf{T}_b(i,j) + (\frac{\sqrt 3}{24} + \frac{a}{2})\mathbf{T}_b(i,j + 1) + (\frac{\sqrt 3}{24} + \frac{a}{2})\mathbf{T}_b(i + 1,j + 1).
\end{equation}
For Type II, the percentages of the hexagonal elements spanned by the green area (top left corner), red area (bottom left corner) and yellow area (right) are $1/2 - \sqrt 3 / 24 - a/2$, $1/2 - \sqrt 3 / 24 - a/2$ and $\sqrt 3 / 12 + a$, respectively. Then, the transmittance of the $(i,j)$th pixel of $\mathbf{T}_g$ can be calculated as 
\begin{equation}\label{e3-4}
\mathbf{T}_g(i,j) = (\frac{1}{2} - \frac{\sqrt 3}{24} -\frac{a}{2})\mathbf{T}_b(i,j) + (\frac{1}{2} - \frac{\sqrt 3 }{24} - \frac{a}{2})\mathbf{T}_b(i+1,j) + (\frac{\sqrt 3}{12} + a)\mathbf{T}_b(i,j + 1).
\end{equation}
\subsection{Theoretical proof of the blue noise ${\rm CA}_{\hexagon}$} 
\label{sec3.2}
In this section, the properties of the optimal ${\rm CA}_{\hexagon}$ are proved based on the RIP of CASSI system. It will be shown that the optimal coded aperture pattern of ${\rm CA}_{\hexagon}$ should obey the blue noise distribution. We also prove the superiority of the blue noise ${\rm CA}_{\hexagon}$ over the traditional random ${\rm CA}_{\square}$ and blue noise ${\rm CA}_{\square}$ in a statistical sense. Second, the optimal ${\rm CA}_{\hexagon}$ used in different snapshots should be complementary to each other.

\subsubsection{Optimality of blue noise sampling }  
According to Eq.~(\ref{e2-6}), ${\bf{A}} = {\bf{H\Psi }} \in {R^{Q_1 \times Q_2}}$ is the sensing matrix, where $Q_1=KV$ ($V=N(M+L-1)$), $ Q_2=NML$. The RIP condition of the sensing matrix $\mathbf{A}$ is critical to design the coded apertures as it guarantees the probability of accurate reconstruction of spectral images\cite{Restricted}. 
Assume $S$ is the number of non-zero elements in the sparse coefficient vector $\bm{\theta}$ of the original signal. The RIP constant ${\delta _s}$ is defined as the smallest constant such that
$(1 - {\delta _s})\left\| {\mathop{\bm{\theta}}\nolimits}  \right\|_2^2 \le {\left\| {{\bf{A}}{\mathop{\bm{\theta}}\nolimits} } \right\|^2} \le (1 + {\delta _s})\left\| {\mathop{\bm{\theta}}\nolimits}  \right\|_2^2$ holds for all $S$-sparse vectors $\bm{\theta}$.
The constant ${\delta _s}$ can be written as\cite{Restricted, classification }
\begin{eqnarray}\label{e7-2}
{\delta _s} = \mathop {\max }\limits_{\tau \subset \left[ {Q_2} \right],\left| \tau \right| \le S} \lambda_{max} \sqrt{({{\bf{A}}_{\left| \tau \right|\left| \tau \right|}} - {\bf{I}})},
\end{eqnarray}
where ${{\bf{A}}_{\left| \tau \right|\left| \tau \right|}} = {\bf{A}}_\tau^T{{\bf{A}}_{ \tau }}$, ${{\bf{A}}_{ \tau}}$ is a matrix whose columns are equal to ${\left| \mathit{\tau} \right|}$ columns of $\mathbf{A}$ indexed by the set $\mathbf{\Omega}$, and ${\lambda _{max} }( \cdot )$ denotes the largest eigenvalue of the argument\cite{structured random matrices}. Let ${\mathbf{\Psi} _{ij}}$ be the $(i,j)$th entry of the basis $\mathbf{\Psi}  \in {R^{ Q_2 \times  Q_2}}$.
Given that ${\bf{A}} = {\bf{H}}\mathbf{\Psi}$, then using the structure of the matrices ${\bf{H}}$, the entries of ${{\bf{A}}_\tau}$ can be expressed as the product of the rows of $\mathbf{H}$ and the columns of $\mathbf{\Psi} $ indexed by the set $\mathbf{\Omega}$:
\begin{eqnarray}\label{e7-4}
{({{\bf{A}}_\tau})_{ij}} = \sum\limits_{r = 0}^{L - 1} {{{({{\mathop{\bm{t}}\nolimits} ^k})}_{i -kV- r{N}}}{\mathbf{\Psi} _{i + r{N'},{\Omega _j}}}},
\end{eqnarray}
where $i = 0, \ldots , Q_1 - 1$, $j = 0, \ldots , {\left| \mathit{\tau} \right|} - 1$, $\bm{t} ^k$ is the vector representation of the coded aperture ${{\bf{T}}^k}$, $k = \left\lfloor {\frac{i}{V}} \right\rfloor $, and $N' = {N^2} - N$. In the above equation, ${\Omega} _j$ is the $j$th element of the set $\mathbf{\Omega}$ and ${\Omega} _j \in \{ 0,...,Q_2 - 1\} $, which means ${\Omega} _j$ is the index selected from the $Q_2$ columns of ${\mathbf{\Psi}}$. Based on Eq.~(\ref{e7-4}), the entries of ${{\bf{A}}_{\left| \tau \right|\left| \tau \right|}}$ are denoted by ${({{\bf{A}}_{\left| \tau  \right|\left| \tau  \right|}})_{ij}}$, which can be calculated as
\begin{eqnarray}\label{e7-5}
 {({{\bf{A}}_{\left| \tau  \right|\left| \tau  \right|}})_{ij}}\!\!=\!\!\sum\limits_{p = 0}^{KV - 1} {{{{\bf{(A}}_\tau ^T)}_{ip}}{{{\rm{(}}{{\bf{A}}_\tau })}_{pj}}} \!\!=\!\!\sum\limits_{k = 0}^{K - 1} {\sum\limits_{p = 0}^{KV - 1} {\sum\limits_{r = 0}^{L - 1} {\sum\limits_{u = 0}^{L - 1} {{{({{\mathop{\bm t}\nolimits} ^k})}_{p \!-\!kV\!-\! rN}}{{({{\mathop{\bm t}\nolimits} ^k})}_{p \!-\!kV\!-\! u{N}}}{\mathbf{\Psi} _{p\! + \!rN',{\Omega _i}}}} } } } {\mathbf{\Psi} _{p \!+\! uN',{\Omega _j}}},
\end{eqnarray} 
for $i,j = 0,...{\left| \mathit{\tau} \right|} - 1$ and $i \ne j$. 
A necessary condition for the RIP is that the diagonal elements of ${{\bf{A}}_{\left| \tau \right|\left| \tau \right|}}$ satisfy $E\left( {({{\bf{A}}_{\left| \tau  \right|\left| \tau  \right|}})_{jj}}\right) = 1$ for all 
$j = 0, \ldots ,S - 1$, where ${\rm{E}}( \cdot )$ means the mathematical expectation\cite{classification,colored,Spatiotemporal_blue_noise }. The matrix ${{\bf{A}}_{\left| \tau \right|\left| \tau \right|}}$ can be normalized by constraining the coded apertures to satisfy $\sum\limits_{k = 0}^{K - 1} {({\bm{t}^k})_{_{p -kV- rN}}^2}  = C$, where C is a selectable constant. Then, the normalized ${{\bf{A}}_{\left| \tau \right|\left| \tau \right|}}$ is defined as ${\bf{B}_{\left| \tau \right|\left| \tau \right|}} = {{\bf{A}}_{\left| \tau \right|\left| \tau \right|}}/{\rm{C}}$. Based on Eq.~(\ref{e7-5}), the elements in ${\bf{B}_{\left| \tau \right|\left| \tau \right|}}$ can be written as
\begin{eqnarray}\label{e7-6}
{({{\bf{B}}_{\left| \tau  \right|\left| \tau  \right|}})_{ij}} = \frac{{\rm{1}}}{{\rm{C}}}\sum\limits_{p = 0}^{V - 1} {\sum\limits_{r = 0}^{L - 1} {\sum\limits_{u = 0}^{L - 1} {{r_{p,r,u}}} } } {\mathbf{\Phi} _{p,r,u}},
\end{eqnarray}
where ${r_{p,r,u}} = \sum\limits_{k = 0}^{K - 1} {({\bm{t}^k}} {)_{p -kV- rN}}{({\bm{t}^k})_{p -kV- uN}}$, ${\mathbf{\Phi} _{p,r,u}} = {\mathbf{\Psi} _{p + rN',{\Omega _i}}}{\mathbf{\Psi} _{p + uN',{\Omega _j}}}$. 
It is assumed that the element of ${\mathbf{\Psi}}$ is bounded such that $\left| {{\mathbf{\Phi} _{p,r,u}}} \right| < {C_1}$ for all ${p,r}$ and ${u}$. Then, Eq.~(\ref{e7-6}) can be rewritten as ${({{\bf{B}}_{\left| \tau  \right|\left| \tau  \right|}})_{ij}}  \le \frac{C_1}{C}\sum\limits_{p = 0}^{V - 1} {\sum\limits_{r = 0}^{L - 1} {\sum\limits_{u = 0}^{L - 1} {{r_{p,r,u}}} } } $. It is noted that ${({{\bf{B}}_{\left| \tau  \right|\left| \tau  \right|}})_{ij}}$ is the sum of the bounded random variables and can be modelled as a sub-Gaussian random variable ${({{\bf{B}}_{\left| \tau  \right|\left| \tau  \right|}})_{ij}} \sim Sub({\alpha ^2})$, where 
\begin{eqnarray}\label{e7-6.51}
\alpha  = \mathop {\max }\limits_{j,k} \mathop  \frac{C_1}{C}\sum\limits_{ p= 0}^{V - 1} {\sum\limits_{r = 0}^{L - 1} {\sum\limits_{u = 0}^{L - 1} {{r_{p,r,u}}} } }.
\end{eqnarray}
Previous works on the RIP for sub-Gaussian variables have established the probability
 \begin{eqnarray}\label{e7-6.52}
 P_r(\left| {B_{\left| \tau  \right|\left| \tau  \right|}  - I} \right| \le {\delta _s}) \ge 1 - \varepsilon,
 \end{eqnarray}
 where $\varepsilon  = 2{(1 + 2/\rho )^2}{e^{ - {\delta _s}(2 - {{(1 + \rho )}^2}KV{C_2}/\alpha }}$ with $\rho  = 2/({e^3} - 1)$, and $C_2$ is a constant independent of $K$ and $V$ \cite{ Spatiotemporal_blue_noise, Nonuniform,colored}.
The probability of accurate reconstruction is $1-\varepsilon $, which can be maximized by increasing the number of measurements $K$ or designing the coded aperture to minimize the parameter $\alpha $ that is related to the variable ${r_{p,r,u}}$. 

Consider the case of ${\rm CA}_{\hexagon}$.
\begin{figure}[!h]
	\centering\includegraphics[width=8cm]{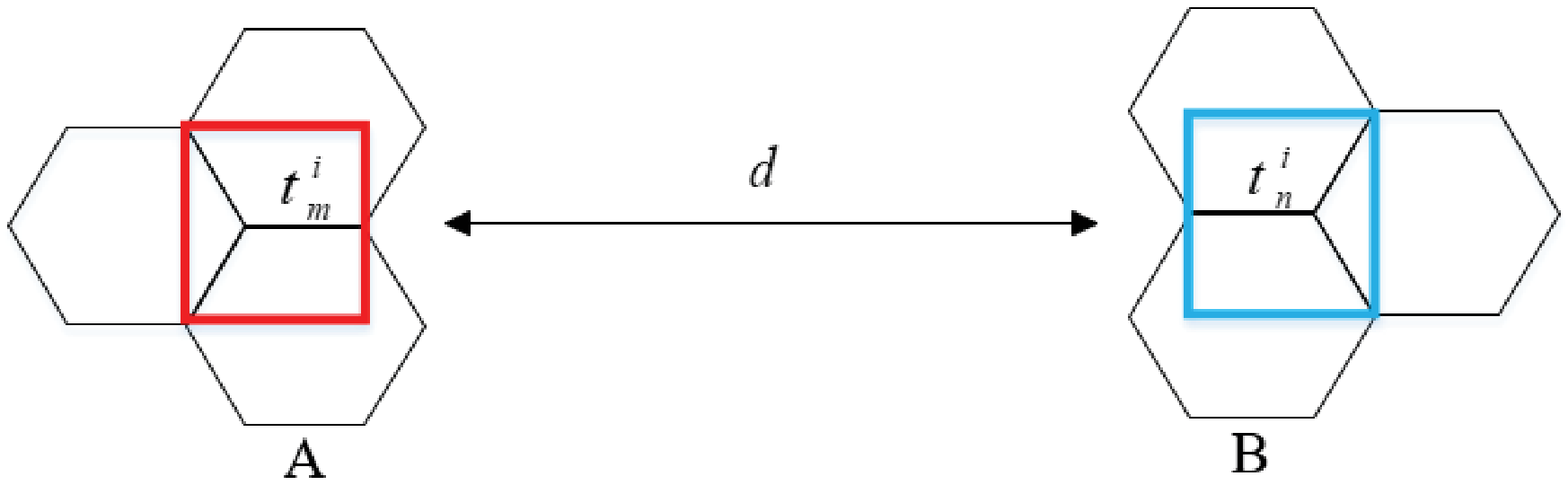} \caption{Example of two macro pixels on the ${\rm CA}_{\hexagon}$.}
	\label{figure11}
\end{figure}
As shown in Fig.~\ref{figure11}, ${t_m^i}$ and ${t_n^i}$ represent the transmittances of two squared tiled elements overlapping the ${\rm CA}_{\hexagon}$. They are modulated by three adjacent hexagonal elements, respectively. Suppose the three adjacent elements associated with ${t_n^i}$ are $t_{n_1}^i$, $t_{n_2}^i$, and $t_{n_3}^i$, thus $t_n^i = a_1t_{n_1}^i + a_2t_{n_2}^i + a_3t_{n_3}^i$, where $a_1$, $a_2$, $a_3$ are the weights corresponding to the overlapped areas. The three adjacent elements associated with ${t_m^i}$ are $t_{m_1}^i$, $t_{m_2}^i$, and $t_{m_3}^i$, thus $t_m^i = b_1t_{m_1}^i + b_2t_{m_2}^i + b_3t_{m_3}^i$, where $b_1$, $b_2$, $b_3$ are the weights corresponding to the overlapped areas. Then, the variable ${r_{p,r,u}}$ can be written as
\begin{eqnarray}\label{e7-6.5}
{r_{p,r,u}} &=& \sum\limits_{i = 0}^{K - 1} {t_n^i} t_m^i = \sum\limits_{i = 0}^{K - 1} {({a_1}t_{{n_1}}^i}  + {a_2}t_{{n_2}}^i + {a_3}t_{{n_3}}^i)({b_1}t_{{m_1}}^i + {b_2}t_{{m_2}}^i + {b_3}t_{{m_3}}^i)\nonumber\\
&=& \sum\limits_{i = 0}^{K - 1}( {{a_1}{b_1}t_{{n_1}}^i} t_{{m_1}}^i + {a_1}{b_2}t_{{n_1}}^it_{{m_2}}^i + {a_1}{b_3}t_{{n_1}}^it_{{m_3}}^i+ {a_2}{b_1}t_{{n_2}}^it_{{m_1}}^i \nonumber\\
&&+ {a_2}{b_2}t_{{n_2}}^it_{{m_2}}^i \!\!+\!\! {a_2}{b_3}t_{{n_2}}^it_{{m_3}}^i\! \!+\!\! {a_3}{b_1}t_{{n_3}}^it_{{m_1}}^i \!\!+\!\! {a_3}{b_2}t_{{n_3}}^it_{{m_2}}^i \!\!+\!\! {a_3}{b_3}t_{{n_3}}^it_{{m_3}}^i).
\end{eqnarray}  

In order to minimize the parameter $\alpha $, the products ${t_{n_1}^i} t_{m_1}^i$, $t_{n_1}^it_{m_2}^i$, $t_{n_1}^it_{m_3}^i$, 
$t_{n_2}^it_{m_1}^i$, $t_{{n_2}}^it_{{m_2}}^i$, $t_{n_2}^it_{m_3}^i$, $t_{{n_3}}^it_{{m_1}}^i$, $t_{n_3}^it_{m_2}^i$, $t_{n_3}^it_{m_3}^i$ should be minimized within a neighborhood of ${\rm CA}_{\hexagon}$. It can be noted that this minimization is achieved when the one-valued entries of $t_{n_1}^i$, $t_{n_2}^i$, and $t_{n_3}^i$ are separated as far as possible from the one-valued entries of $t_{m_1}^i$, $t_{m_2}^i$, and $t_{m_3}^i$. In addition, according to the first line of Eq.~(\ref{e7-6.5}), the number of ones among $t_{m_1}^i$, $t_{m_2}^i$, $t_{m_3}^i$, $t_{n_1}^i$, $t_{n_2}^i$, and $t_{n_3}^i$ should be as small as possible to minimize the parameter $\alpha $. These conditions mentioned above, in essence, require the hexagonal sampling to satisfy the blue noise distribution\cite{ Spatiotemporal_blue_noise}.

In the above, we have demonstrated that under the hexagonal lattice, the optimal coded aperture should obey the blue noise distribution. It is natural to ask if the blue noise ${\rm CA}_{\hexagon}$ outperforms other kinds of coded apertures under rectangular lattice. The superiority of the blue noise ${\rm CA}_{\hexagon}$ over the random ${\rm CA}_{\square}$ and blue noise ${\rm CA}_{\square}$ is shown next. In a statistical sense, a good coded aperture should minimize the mathematical expectation of the parameter $\alpha $ in Eq.~(\ref{e7-6.51}). This is equivalent to minimizing the mathematical expectation of the variable ${r_{p,r,u}}$. In this paper, we prove that using the blue noise ${\rm CA}_{\Square}$ will reduce the mean value of ${r_{p,r,u}} $ compared to using the random ${\rm CA}_{\Square}$. In addition, using the blue noise ${\rm CA}_{\hexagon}$ will further reduce the mean value of ${r_{p,r,u}} $ compared to using the blue noise ${\rm CA}_{\Square}$, i.e.,
\begin{eqnarray}\label{e7-7}
E(r_{p,r,u})_{SR}}>E(r_{p,r,u})_{SB}>{E(r_{p,r,u})_{HB},
\end{eqnarray}
where $E(r_{p,r,u})_{HB}$, $E(r_{p,r,u})_{SB}$, and $E(r_{p,r,u})_{SR}$ represent the mean values of ${r_{p,r,u}} $ when using the blue noise ${\rm CA}_{\hexagon}$, blue noise ${\rm CA}_{\Square}$, and random ${\rm CA}_{\Square}$, respectively. The derivation of the first inequality in Eq.~(\ref{e7-7}) is provided in Appendix A, and the derivation of the second inequation is given in Appendix B.
\subsubsection{Optimality of complementary coding for multishots}
It can be observed that decreasing $C$ will increase the value of $\alpha$ in Eq.~(\ref{e7-6.51}) \cite{colored}. For the ${\rm CA}_{\hexagon}$, 
\begin{eqnarray}\label{e7-6.6}
C &=& \sum\limits_{i = 0}^{K - 1} {{{(t_n^i)}^2}}  = \sum\limits_{i = 0}^{K - 1} {({a_1}t_{{n_1}}^i}  + {a_2}t_{{n_2}}^i + {a_3}t_{{n_3}}^i{)^2}\nonumber\\
&=&\sum\limits_{i = 0}^{K - 1} {{a_1}^2t{_{n_1}^i}^2}+{{a_2}^2t{_{n_2}^i}^2}+{{a_3}^2t{_{n_3}^i}^2}+2{a_1}{a_2}t_{n_1}^it_{n_2}^i + 2{a_1}{a_3}t_{n_1}^it_{n_3}^i + 2{a_2}{a_3}t_{{n_2}}^it_{n_3}^i. 
\end{eqnarray}
If the ${\rm CA}_{\hexagon}$ satisfies the blue noise distribution, it is natural to assume that two adjacent pixels are unlikely to value one at the same time. Then,
\begin{eqnarray}\label{e7-6.7}
C =\sum\limits_{i = 0}^{K - 1} {{a_1}^2t{_{n_1}^i}^2}+{{a_2}^2t{_{n_2}^i}^2}+{{a_3}^2t{_{n_3}^i}^2}.
\end{eqnarray}
It is noted that $\sum\limits_{i = 0}^{K - 1} {t{{_{{n_1}}^i}^2}}  \ge {\rm{1}}$, $\sum\limits_{i = 0}^{K - 1} {t{{_{{n_2}}^i}^2}}  \ge {\rm{1}}$, and $\sum\limits_{i = 0}^{K - 1} {t{{_{{n_3}}^i}^3}}  \ge {\rm{1}}$. In order to minimize $C$, we should have $\sum\limits_{i = 0}^{K - 1} {t{{_{{n_1}}^i}^2}}=\sum\limits_{i = 0}^{K - 1} {t{{_{{n_2}}^i}^2}}=\sum\limits_{i = 0}^{K - 1} {t{{_{{n_3}}^i}^2}}=1$. It is equivalent to design the ${\rm CA}_{\hexagon}$ such that the pixels are complementary among multishots in every spatial position on the coded apertures.

\subsection{Generation of blue noise ${\rm CA}_{\hexagon}$}
\label{sec3.3}
\begin{figure}[!h]
	\centering\includegraphics[width=12cm]{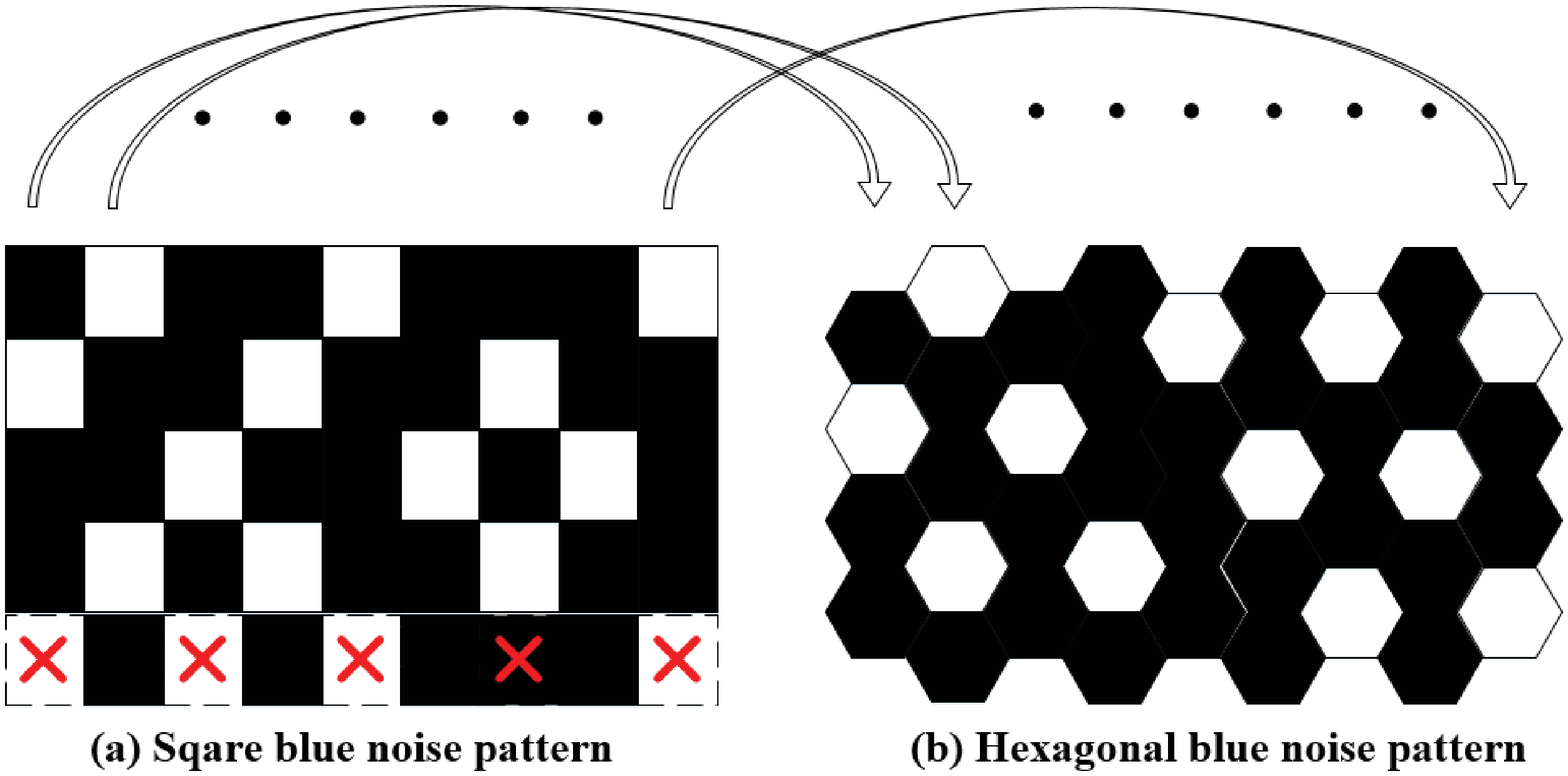} \caption{The method to generate the blue noise ${\rm CA}_{\hexagon}$.}
	\label{figure3}
\end{figure}

This section describes the method to generate blue noise ${\rm CA}_{\hexagon}$. Suppose the size of target is $N \times M$ in the spatial domain. The method to generate the blue noise ${\rm CA}_{\hexagon}$ is shown in Fig.~\ref{figure3}. We first use the method proposed in \cite{blue_noise_square} to generate an $(N+1) \times (M+1)$ blue noise pattern with square pixels as illustrated in Fig.~\ref{figure3}(a). Then, the square coded aperture in Fig.~\ref{figure3}(a) is transformed to the blue noise ${\rm CA}_{\hexagon}$ pattern in Fig.~\ref{figure3}(b) by fitting the pixels column by column. It is shown that the odd column of the blue noise ${\rm CA}_{\hexagon}$ includes $N$ elements, while the even column includes $N+1$ elements. Thus, for the square blue noise pattern in Fig.~\ref{figure3}(a), we need to withdraw the last pixels in the odd columns as marked by the red crosses. 

Using the matrix $\mathbf{T}_g$ defined in Eqs.~(\ref{e3-1})-(\ref{e3-4}), the $\mathbf{H}$ matrix of the system in Eq.~(\ref{e2-4}) can be constructed. 
Figure~\ref{figure2} compares the $\mathbf{H}$ matrices based on the random ${\rm CA}_{\Square}$ and blue noise ${\rm CA}_{\hexagon}$ for $K=2$, $N=M=6$, and $L=3$. The transmittance of the coded apertures is 50\%, and the offset ratio $a$ between the square detector array and blue noise ${\rm CA}_{\hexagon}$ is 0. It is important to remark that the dimension of $\mathbf{H}$ is large in practice, and the example in Fig.~\ref{figure2} is only for illustrative purposes. In Fig.~\ref{figure2} (a), the entries of the ${\rm CA}_{\Square}$ are all binary, ${\bf{T}}_{ij}^k \in {\rm{\{ 0}},{\rm{1\} }} $, following a Bernoulli distribution. As can be seen from Fig.~\ref{figure2} (a), there exist several zero-valued elements in succession. Thus, the sampling is nonuniform, leading to the loss of structure information during the data acquisition stage. On the other hand, the non-zero entries in Fig.~\ref{figure2} (b) are distributed more uniformly, which is beneficial to capture more information from the target. In addition, the entry values in Fig.~\ref{figure2} (b) vary in the interval [0,1], which increases the degrees of freedom in the sensing matrix. In practice, the grey-scale entry values can be modified by changing the offset ratio $a$ in Fig.~\ref{figure5}.

\section{Simulation results}
\label{sec4}
\begin{figure}[!h]
	\centering\includegraphics[width=13.5cm]{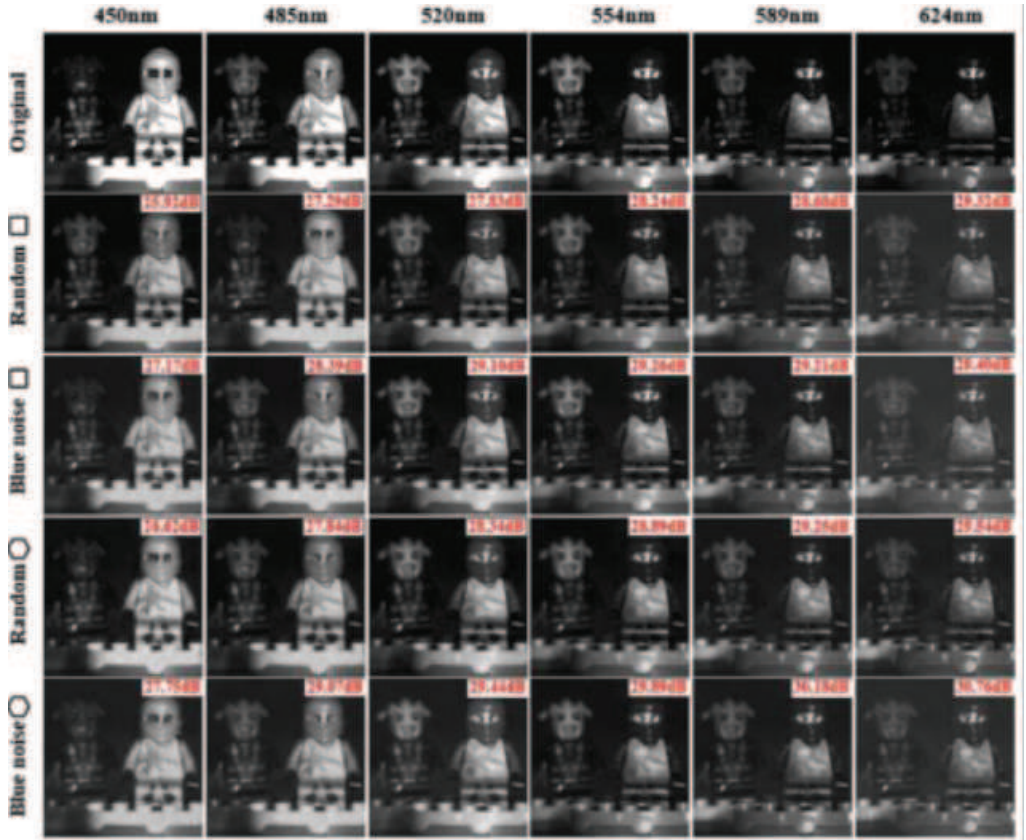} \caption{Simulations of the hyperspectral imaging methods using two snapshots and complementary coded apertures with 50\% transmittance. From top to bottom, it respectively shows the original spectral images (first row), the reconstructed images using the random ${\rm CA}_{\Square}$ (second row), blue noise ${\rm CA}_{\Square}$ (third row), random ${\rm CA}_{\hexagon}$ (fourth row), and blue noise ${\rm CA}_{\hexagon}$ (fifth row).}
	\label{figure6}
\end{figure}
\begin{figure}[!h]
	\centering\includegraphics[width=13.5cm]{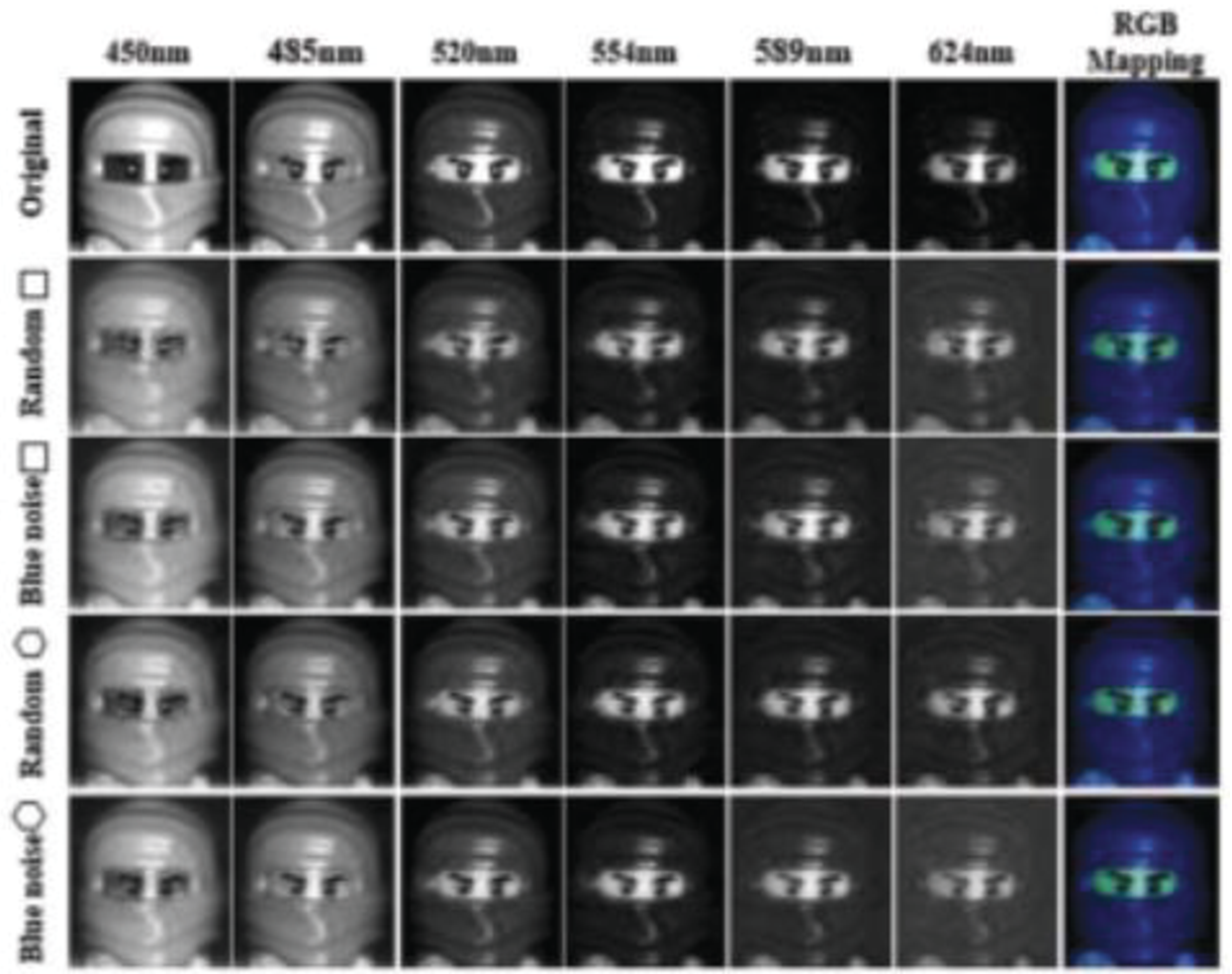} \caption{Magnified details of the reconstructed images using two snapshots and complementary coded apertures with 50\% transmittance. From top to bottom, it respectively shows the magnified regions on the original spectral images (first row), the reconstructed images using the random ${\rm CA}_{\Square}$ (second row), blue noise ${\rm CA}_{\Square}$ (third row), random ${\rm CA}_{\hexagon}$ (fourth row), and blue noise ${\rm CA}_{\hexagon}$ (fifth row). The average PSNRs across the six spectral bands are 27.88 dB, 28.76 dB, 28.41 dB, and 29.52 dB, respectively.}
	\label{figure6.5}
\end{figure} 
\begin{figure}[!h]
	\centering\includegraphics[width=12cm]{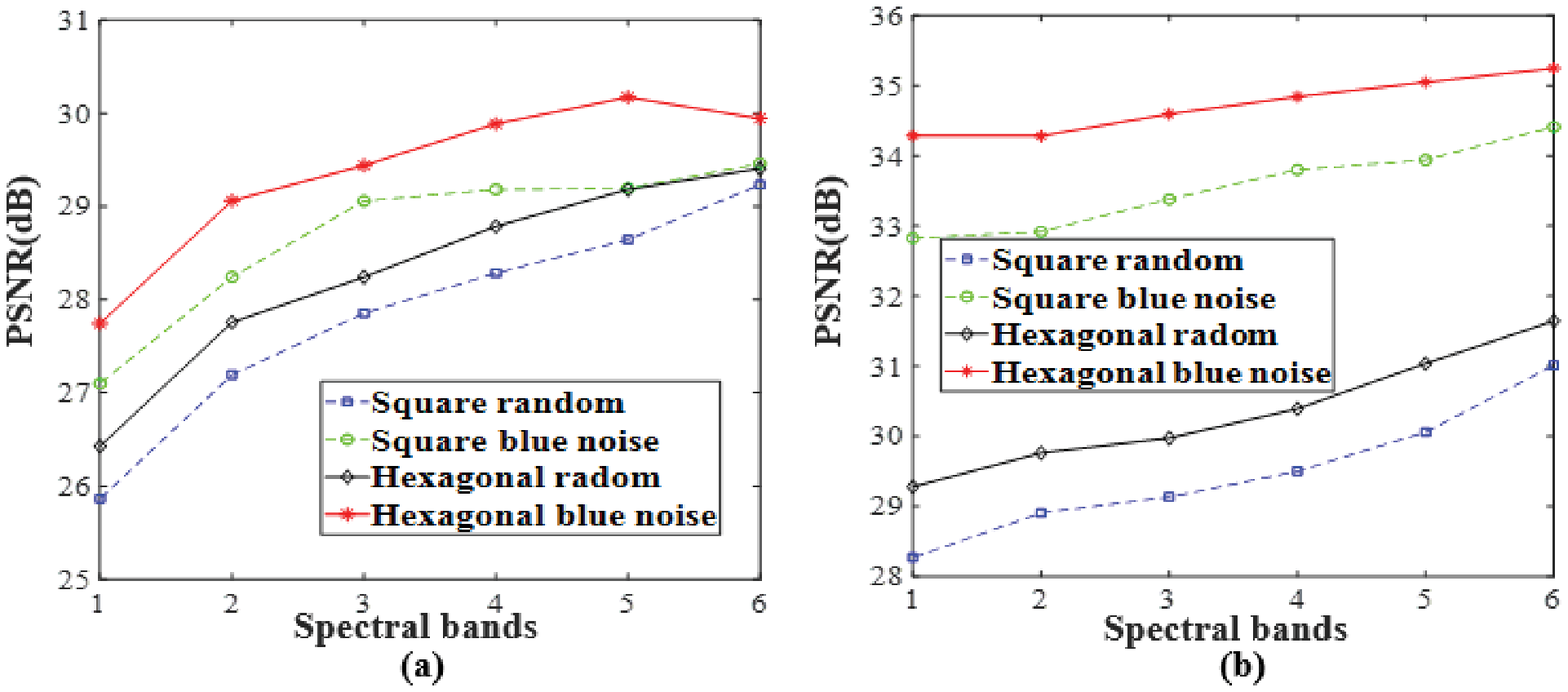} \caption{The comparison of the average reconstruction PSNRs in different spectral bands obtained by random ${\rm CA}_{\Square}$ (blue dash-square curve), blue noise ${\rm CA}_{\Square}$ (green dash-circle curve), random ${\rm CA}_{\hexagon}$ (black dash-prismatic curve), and blue noise ${\rm CA}_{\hexagon}$ (red dash-star curve). The transmittance of the coded apertures is 50\% in (a) and 25\% in (b).}
	\label{figure7}
\end{figure}
This section provides a set of simulations to assess the reconstruction performance obtained by the proposed hexagonal blue noise coding strategy. In addition, the performance of blue noise ${\rm CA}_{\hexagon}$ is compared with that of random ${\rm CA}_{\Square}$, blue noise ${\rm CA}_{\Square}$, and random ${\rm CA}_{\hexagon}$. The spatial dimension of the original spectral data cube $\mathbf{F}$ is $N=M=256$, and the spectral dimension is $L=6$. The wavelengths of the 6 spectral bands are 450nm, 485nm, 520nm, 554nm, 589nm and 624nm, respectively. The spatial dimension of the FPA detector is ${256\times(256+6-1)=256\times261}$, and the resolution of the coded aperture is the same as that of the detector. The GPSR algorithm is used to recover the spectral images from the compressive measurements.

\begin{figure}[!h]
	\centering\includegraphics[width=13.5cm]{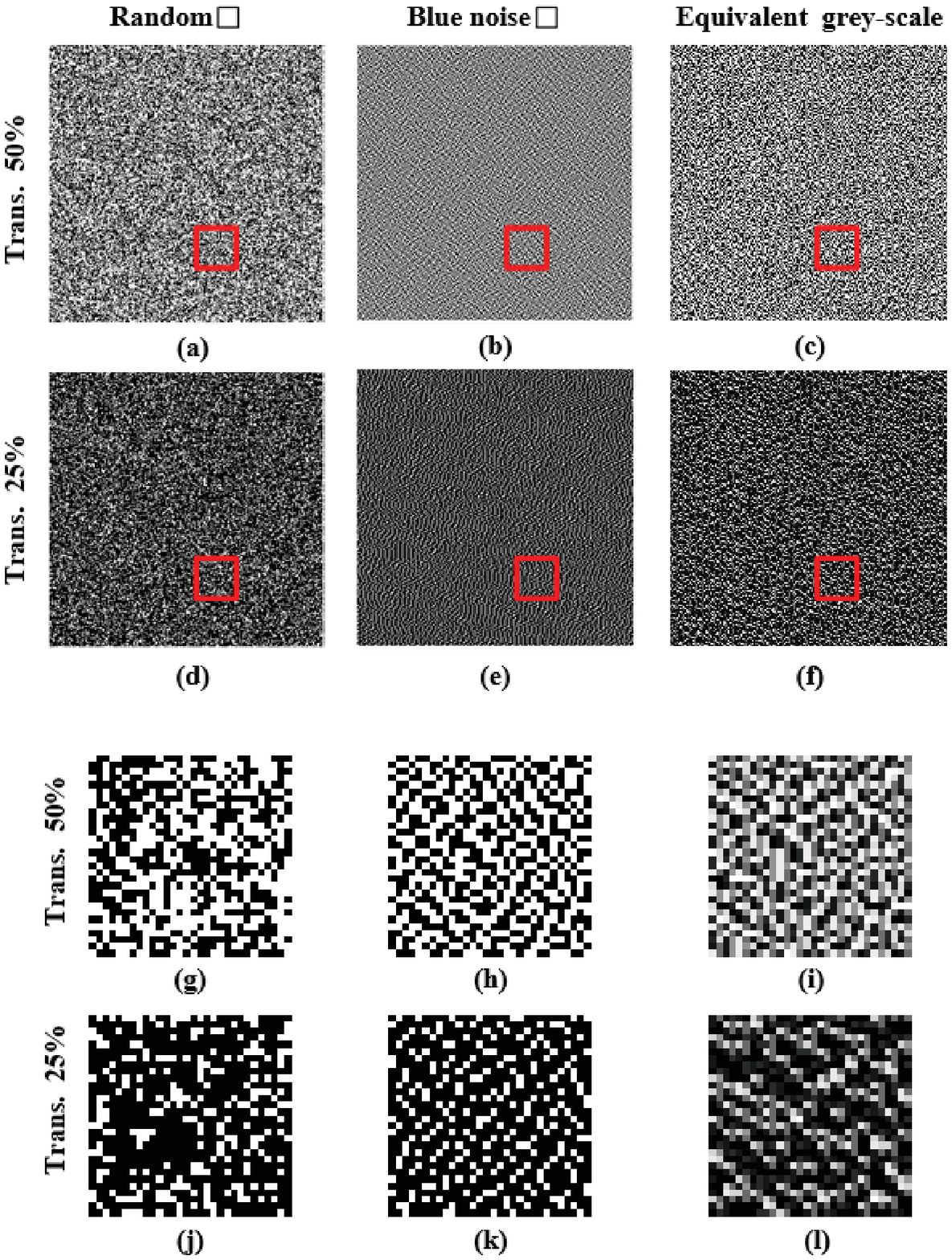} \caption{Examples of different coded apertures. (a)-(f) are examples of the random ${\rm CA}_{\Square}$, blue noise ${\rm CA}_{\Square}$, and equivalent grey-scale modulation patterns corresponding to the blue noise ${\rm CA}_{\hexagon}$. (g)-(l) illustrate the magnified regions in the coded apertures surrouned by the red boxes. The transmittance of the coded apertures is 50\% in the first and third rows. The transmittance of the coded apertures is 25\% in the second and fourth rows.}
	\label{figure8}
\end{figure}
Figure~\ref{figure6} shows the simulation results based on the coded apertures with 50\% transmittance, and the number of snapshots is $K=2$.
The target under detection includes two Lego\textregistered figures. The first row in Fig.~\ref{figure6} shows the original spectral images in the 6 spectral bands. The second row illustrates the reconstructed spectral images using the random ${\rm CA}_{\Square}$. The third row illustrates the reconstruction results based on the blue noise ${\rm CA}_{\Square}$, where the coded apertures constitute squared tiled elements, but the block/unblock pattern obeys the blue noise sampling distribution. The fourth row illustrates the reconstruction results using random ${\rm CA}_{\hexagon}$. The fifth row illustrates the reconstruction results using the proposed blue noise ${\rm CA}_{\hexagon}$. It is noted that the weight coefficient of the sparsity regularization in GPSR algorithm will influence the reconstruction performance\cite{GPSR}. In these simulations, we use the line search method to find out the best weight coefficients for all cases. From the second row to the fifth row, the weight coefficients are set to be 2e-4, 8e-5, 8e-5 and 4e-5, respectively. The peak signal to noise ratios (PSNR) of all reconstructed spectral images are presented as well. Figure~\ref{figure6.5} shows the magnified details of the heads of Lego features in Fig.~\ref{figure6}. 

Figure~\ref{figure7}(a) compares the average PSNRs of the reconstructed images in different spectral bands. 
In Fig.~\ref{figure7}(a), the $x$-axis indicates the order numbers of the spectral bands, and the $y$-axis shows the average reconstruction PSNRs. For each simulation setting, we run the reconstruction algorithm for 5 times with different coded apertures and calculate the average PSNRs in all spectral bands.  
It is observed from Fig.~\ref{figure6}, Fig.~\ref{figure6.5} and Fig.~\ref{figure7}(a) that the blue noise ${\rm CA}_{\Square}$ lead to superior reconstruction quality over the random ${\rm CA}_{\Square}$ in all spectral bands. That is because the blue noise coded apertures sample the images more uniformly, so that can extract more structure characteristics of the underlying spectral scene. In addition, the proposed blue noise ${\rm CA}_{\hexagon}$ can further improve the reconstruction performance compared to the blue noise ${\rm CA}_{\Square}$. That is because the geometric dislocation between the hexagonal coded apertures and the square detector array introduces an equivalent grey-scale spatial modulation on the spectral images, thus increasing the degrees of freedom in the sensing matrix.   

\begin{figure}[!h]
	\centering\includegraphics[width=12cm]{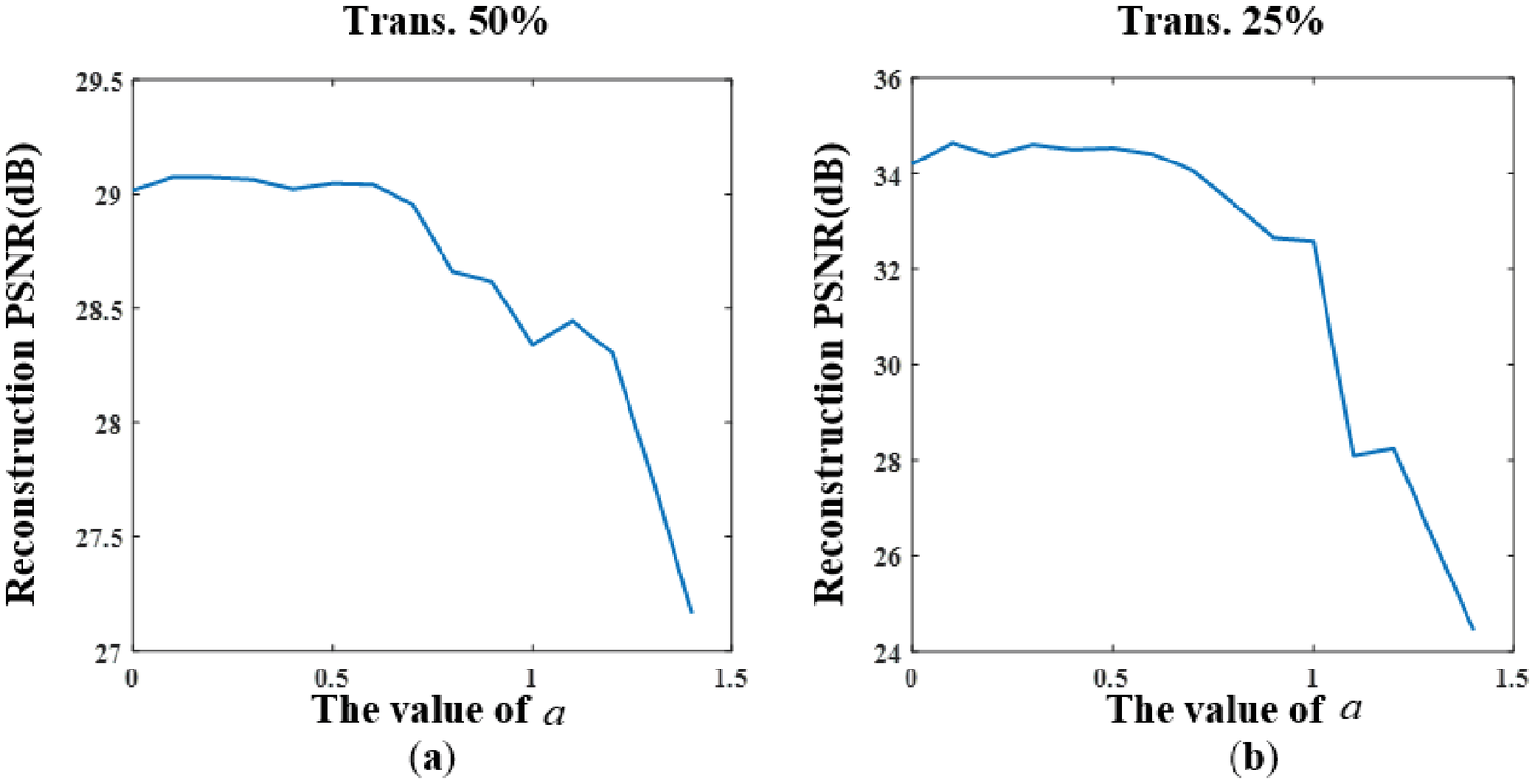} \caption{Average reconstruction PSNR using two snapshots with respect to the offset between the square detector array and blue noise ${\rm CA}_{\hexagon}$, where the transmittance of the coded apertures is 50\% in (a) and 25\% in (b).}
	\label{figure9}
\end{figure}
Figures~\ref{figure8}(a)-\ref{figure8}(c) illustrate the examples of the random ${\rm CA}_{\Square}$, blue noise ${\rm CA}_{\Square}$, and the equivalent grey-scale modulation pattern corresponding to the blue noise ${\rm CA}_{\hexagon}$ with 50\% transmittance. Figures~\ref{figure8}(g)-\ref{figure8}(i) magnify the small regions within the red boxes in Figs.~\ref{figure8}(a)-\ref{figure8}(c). It is observed that the sampling of blue noise ${\rm CA}_{\Square}$ is more uniform than that of the random ${\rm CA}_{\Square}$. In addition, the equivalent grey-scale modulation pattern of the blue noise ${\rm CA}_{\hexagon}$ distributes more uniformly and provides more degrees of freedom on the elements being coded.
Let's consider a more general case, where the square detector array is moved along the $x$-axis with respect to the ${\rm CA}_{\hexagon}$, as shown in Fig.~\ref{figure5}. The offset between the square array and the ${\rm CA}_{\hexagon}$ is $aL$. The reconstruction performance will change with the offset ratio $a$, as shown in Fig.~\ref{figure9}(a). The best value of $a$ is in the range of $(0,0.6)$.
\begin{figure}[!h]
	\centering\includegraphics[width=13.5cm]{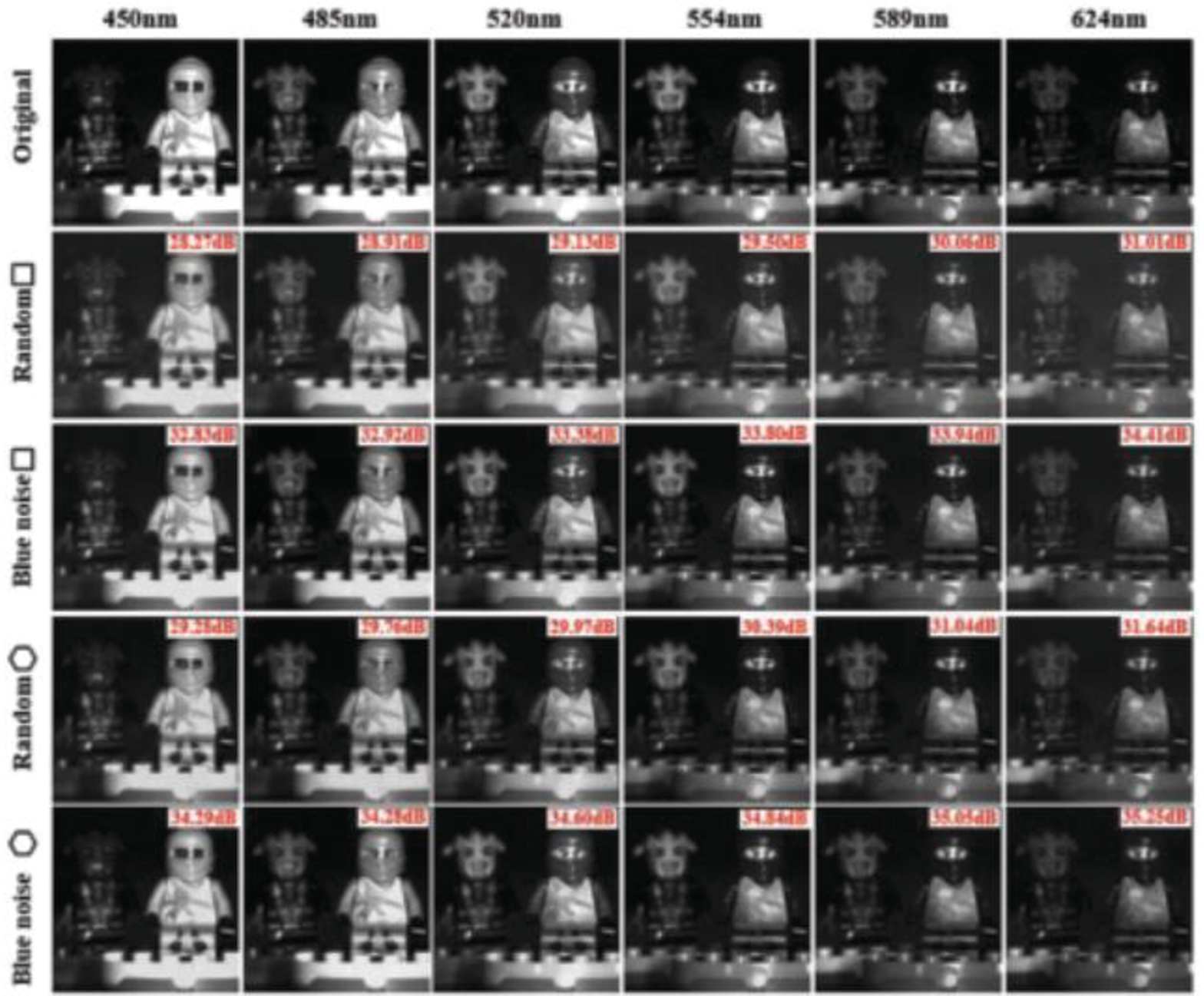} \caption{Simulations of the hyperspectral imaging methods using four snapshots and complementary coded apertures with 25\% transmittance. From top to bottom, it respectively shows the original spectral images (first row), the reconstructed images using the random ${\rm CA}_{\Square}$ (second row), blue noise ${\rm CA}_{\Square}$ (third row), random ${\rm CA}_{\hexagon}$ (fourth row), and blue noise ${\rm CA}_{\hexagon}$ (fifth row).}
	\label{figure10}
\end{figure}
\begin{figure}[!h]
	\centering\includegraphics[width=13.5cm]{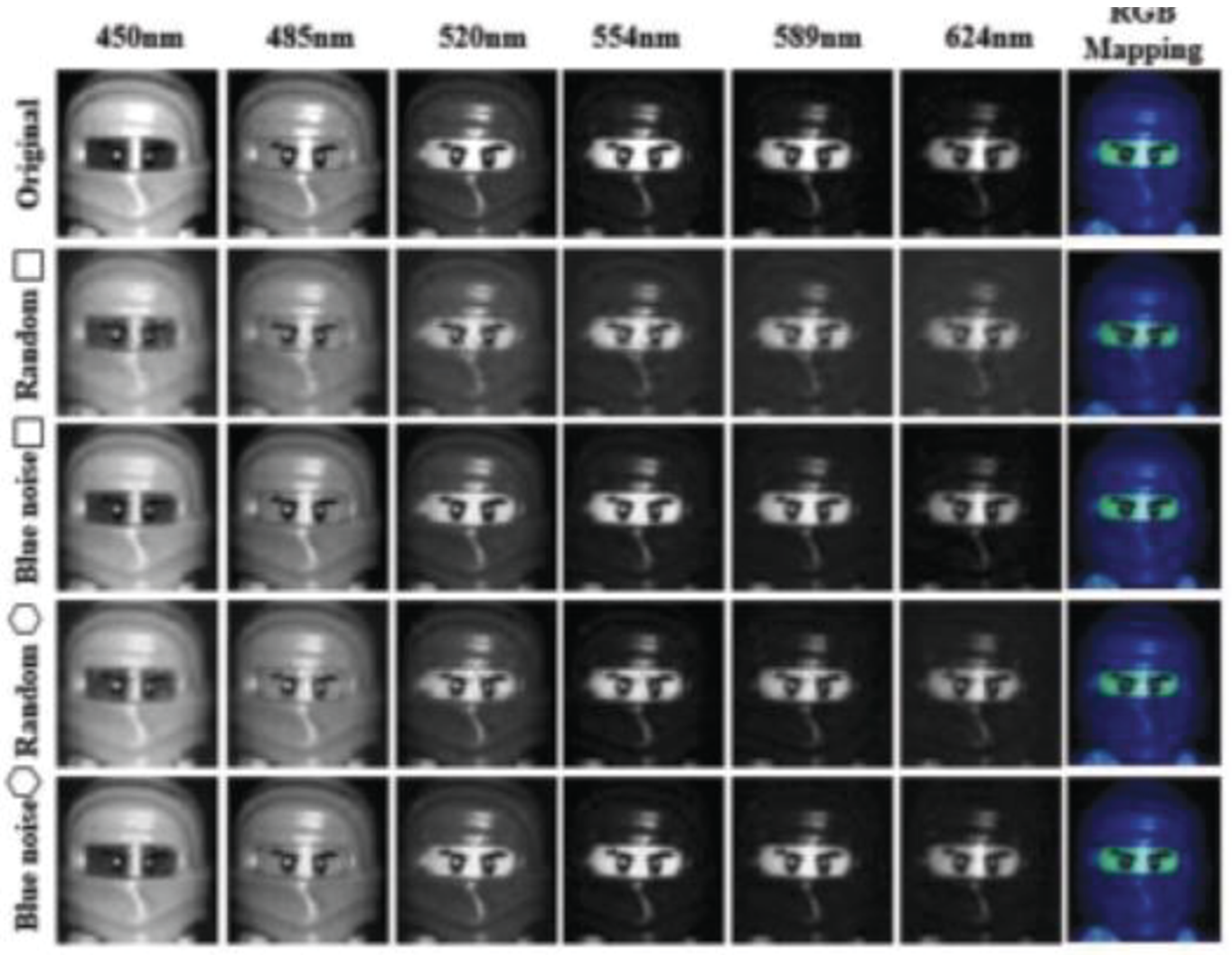} \caption{Magnified details of the reconstructed images using four snapshots and complementary coded apertures with 25\% transmittance. From top to bottom, it respectively shows the magnified regions on the original spectral images (first row), the reconstructed images using the random ${\rm CA}_{\Square}$ (second row), blue noise ${\rm CA}_{\Square}$ (third row), random ${\rm CA}_{\hexagon}$ (fourth row), and blue noise ${\rm CA}_{\hexagon}$ (fifth row). The average PSNRs across the six spectral bands are 29.48 dB, 33.55 dB, 30.35 dB, and 34.72 dB, respectively.}
	\label{figure10.5}
\end{figure}

Figure~\ref{figure10} shows the simulation results based on the coded apertures with 25\% transmittance, and the number of snapshots is $K=4$. From the second row to the fifth row, it shows the reconstructed spectral images using the random ${\rm CA}_{\Square}$, blue noise ${\rm CA}_{\Square}$, random ${\rm CA}_{\hexagon}$ and blue noise ${\rm CA}_{\hexagon}$, respectively. In these simulations, we use the line search method to find out the best weight coefficients for all cases. From the second row to the fifth row, the weight coefficients are set to be 2e-4, 1e-4, 6e-5 and 1e-5, respectively. Figure~\ref{figure10.5} shows the magnified details in Fig.~\ref{figure10} for all simulations. 
Figure~\ref{figure7}(b) compares the average reconstruction PSNRs in different spectral bands obtained by running the simulations for 5 times. The order of the reconstruction accuracy from lower to higher is the random ${\rm CA}_{\Square}$, random ${\rm CA}_{\hexagon}$, blue noise ${\rm CA}_{\Square}$, and blue noise ${\rm CA}_{\hexagon}$.
Figures~\ref{figure8}(d)-\ref{figure8}(f) illustrate the examples of the random ${\rm CA}_{\Square}$, blue noise ${\rm CA}_{\Square}$, and the equivalent grey-scale modulation pattern corresponding to the blue noise ${\rm CA}_{\hexagon}$ with 25\% transmittance. Figures~\ref{figure8}(j)-\ref{figure8}(l) show the magnified regions in the coded apertures mentioned above.
As the square detector array moves along the x-axis with respect to the ${\rm CA}_{\hexagon}$, the curve of the reconstruction PSNR with respect to the parameter $a$ is shown in Fig.~\ref{figure9}(b). The best value of $a$ is in the range of $(0,0.6)$. Based on all of the simulations above, we can conclude that the proposed hexagonal blue noise coding strategy can effectively improve the quality of the reconstructed spectral images.

\section{Conclusions}
\label{sec5}

This paper proposes a design method of blue noise ${\rm CA}_{\hexagon}$ to improve the reconstruction performance of the CASSI systems. In the imaging process, the spectral data cube is modulated by the blue noise ${\rm CA}_{\hexagon}$, then shifted by a dispersive element, and projected onto the detector. Based on the sparsity assumption, the GPSR algorithm is applied to recover the original spectral images from the compressive measurements. The theoretical proof of the superiority of the blue noise ${\rm CA}_{\hexagon}$ based on the RIP of CS theory is derived. In addition, the impact of the offset between the square detector array and the blue noise ${\rm CA}_{\hexagon}$ is discussed. A set of simulations are provided to assess the proposed coding strategy, and compare it to the random ${\rm CA}_{\Square}$, blue noise ${\rm CA}_{\Square}$, and random ${\rm CA}_{\hexagon}$. In the future, we plan to establish a testbed to verify the proposed blue noise ${\rm CA}_{\hexagon}$ method in experiments.  

\appendix
\section{Appendix: the derivation of the first inequation in Eq.~(\ref{e7-7})}
\label{sec7}

First, let's consider the case when using the random ${\rm CA}_{\Square}$. According to the formula of $r_{p,r,u}$ in Section~\ref{sec3.2}, $E(r_{p,r,u})_{SR}$ can be calculated as $E\left(\sum\limits_{i = 0}^{K - 1} {t_{p-kV-rN}^i} t_{p-kV-uN}^i\right)$. Replace the subscripts $p-kV-rN$ and $p-kV-uN$ to $m$ and $n$, respectively. Then, we have
\begin{eqnarray}\label{e7-7.1}
E(r_{p,r,u})_{SR}&=& E\left(\sum\limits_{i = 0}^{K - 1} {t_m^i} t_n^i\right)= \sum\limits_{i = 0}^{K - 1} {E(t_m^it_n^i)}.
\end{eqnarray}
Suppose the mean value of the random ${\rm CA}_{\Square}$ pixels is $g$, where g is equal to the grey level of the blue noise 
coded apertures that will be discussed shortly\cite{blue_noise_square}. For a large area, the grey level of the blue noise coded apertures is approximate to their transmittance. That means the random ${\rm CA}_{\Square}$ has the same transmittance as the blue noise coded apertures. The pixels in random ${\rm CA}_{\Square}$ are considered independent identical random variables, thus Eq.~(\ref{e7-7.1}) becomes 
\begin{eqnarray}\label{e7-7.2}
E(r_{p,r,u})_{SR} &=&  \sum\limits_{i = 0}^{K - 1} {E(t_m^i)} E(t_n^i)
= K{g^2}.
\end{eqnarray}

Next, let's consider the case when using the blue noise ${\rm CA}_{\Square}$, the mathematical expectation of ${r_{p,r,u}} $ can be calculated as
\begin{eqnarray}\label{e7-8}
E(r_{p,r,u})_{SB} \!\! &=&\!\!E\left (\sum\limits_{i = 0}^{K - 1} {t_m^i} t_n^i\right) 
=E(t_m^0t_n^0) + E(t_m^1t_n^1) + ...E(t_m^{K - 1}t_n^{K - 1})\
\nonumber\\
\!\!&=&\!\! 1 \!\cdot \!1 \!\cdot \! P_r(t_m^0 =1;t_n^0 = 1)+1 \!\cdot\!  0 \!\cdot\!  P_r(t_m^0 =1 ;t_n^0 = 0)+0 \!\cdot\!  1 \!\cdot \! P_r(t_m^0 =0;t_n^0 = 1)\nonumber\\
&&+ 0 \cdot  0 \cdot  P_r(t_m^0 =0 ;t_n^0 = 0)+...1 \cdot  1 \cdot  P_r(t_m^{K-1}=1;t_n^{K-1} = 1)\nonumber\\
&&+1 \cdot  0 \cdot  P_r(t_m^{K-1} =1 ;t_n^{K-1} = 0)+ 0 \cdot  1 \cdot  P_r(t_m^{K-1} =0;t_n^{K-1} = 1)\nonumber\\
&&+0 \cdot  0 \cdot  P_r(t_m^{K-1} =0 ;t_n^{K-1} = 0)\nonumber\\
&=& P_r(t_m^0 =1;t_n^0 = 1)+...+P_r(t_m^{K-1} =1;t_n^{K-1} = 1).
\end{eqnarray}
The pixels at the same location of the blue noise ${\rm CA}_{\Square}$ for multishots should have the same probability distribution when taking $K$ snapshots. Thus, we have $P_r(t_m^0 =1;t_n^0 = 1)=...=P_r(t_m^{K-1} =1;t_n^{K-1} = 1)$. 
Then,
\begin{eqnarray}\label{e7-9}
E(r_{p,r,u})_{SB}
=K\cdot P_r(t_m^i = 1;t_n^i = 1)
=K\cdot P_r(t_m^i = 1|t_n^i = 1 )\cdot  P_r(t_n^i = 1), \;\textrm{for}\; \forall i.
\end{eqnarray}

According to Ref.\cite{hexagonal grids,blue_noise_square,multitone dithering}, we make some hypotheses. First,
suppose 
\begin{eqnarray}\label{e7-10.5}
P_r\left \{t_m^i = 1| t_n^i = 1,d < {\lambda _b}\right \} = 0,
\end{eqnarray}
where $d$ is the distance between pixel $t_m^i$ and $t_n^i$, $\lambda _b$ is referred to as the principal wavelength of blue noise\cite{hexagonal grids,Green-noise digital,Digital halftoning,Minimizing stochastic}. That is, when the distance between pixel $t_m^i$ and $t_n^i$ is less than ${\lambda _b}$, the probability of $t_m^i=1$ given $t_n^i=1$ is zero. Second, suppose 
\begin{eqnarray}\label{e7-10}
P_r\left \{t_m^i = 1| t_n^i = 1,d > {\lambda _b}\right \} =
\left\{ \begin{array}{l}
g, \; \textrm{for} \;0 < g \le \frac{1}{2}\\
1 - g,\; \textrm{for}\; \frac{1}{2} < g \le 1.
\end{array} \right.
\end{eqnarray}
That is, when the distance between $t_m^i$ and $t_n^i$ is bigger than ${\lambda _b}$, the probability of $t_m^i=1$ given $t_n^i=1$ is $g$, when $g$ falls in the range of $(0,{1 \over 2}]$, while the probability of $t_m^i=1$ given $t_n^i=1$ is $1-g$, when $g$ falls in the range of $({1 \over 2},1]$. 
Based on these two assumptions mentioned above, Eq.~(\ref{e7-9}) becomes    
\begin{eqnarray}\label{e7-11}
E(r_{p,r,u})_{SB}\!\!\!\!\!&=&\!\!\!\!\!K\!\cdot\!\left \{P_r(t_m^i = 1|t_n^i = 1, d \!>\! \lambda _b)\!\cdot\!  P_r(d \!> \!\lambda _b)
\!\!+\!\!P_r(t_m^i = 1|t_n^i = 1, d\! <\! \lambda _b)\!\cdot\!  P_r(d \!< \!\lambda _b)\right \}\nonumber\\
\!\!&\cdot &\!\! P_r(t_n^i = 1).
\end{eqnarray}
Substituting Eqs.~(\ref{e7-10.5}), and (\ref{e7-10}) to Eq.~(\ref{e7-11}), we have
\begin{eqnarray}\label{e7-11.5}
E(r_{p,r,u})_{SB}&=& K\!\cdot\! P_r(t_m^i= 1|t_n^i = 1,d > {\lambda _b} ) \cdot P_r(d > {\lambda _b})  \cdot P_r(t_n^i= 1).
\end{eqnarray}
According to the definition of $g$, $P_r(t_n^i= 1)=g$, where $g$ is the transmittance of the coded aperture.
When $0 < g \le \frac{1}{2}$,
\begin{eqnarray}\label{e7-13}
E(r_{p,r,u})_{SB} = K\cdot{g^2}  \cdot {P_r}(d > {\lambda _b}) < K{g^2}=E(r_{p,r,u})_{SR}.
\end{eqnarray}
On the other hand, when $\frac{1}{2} < g \le 1$,
\begin{eqnarray}\label{e7-15}
E(r_{p,r,u})_{SB} = K\cdot g(1 - g)  \cdot {P_r}(d > {\lambda _b}) < K\cdot{g^2}{P_r}(d > {\lambda _b}) < K{g^2}=E(r_{p,r,u})_{SR}.
\end{eqnarray}
In summary, we have $E(r_{p,r,u})_{SR}>E(r_{p,r,u})_{SB}$.

\section{Appendix: the derivation of the second inequation in Eq.~(\ref{e7-7})}
\label{sec8} 
Let's consider the case of blue noise ${\rm CA}_{\hexagon}$.
As shown in Fig.~\ref{figure11}, ${t_m^i}$ and ${t_n^i}$ are two square pixels in the equivalent grey-scale modulation pattern of the blue noise ${\rm CA}_{\hexagon}$, and $d$ is the distance between them. The parameters $a$ and $b$ are the values of ${t_m^i}$ and ${t_n^i}$, respectively. Suppose the three adjacent hexagonal elements associated with ${t_m^i}$ constitute a macro pixel $A$, and the three adjacent hexagonal elements associated with ${t_n^i}$ constitute a macro pixel $B$. Suppose the values of hexagonal elements obey the blue noise distribution, and the distance among the three adjacent hexagonal elements in the same macro pixel is smaller than $ \lambda _b$. Then, there will be at most one of the three adjacent hexagonal elements has value 1. 

Similar to Eq.~(\ref{e7-8}), the mathematical expectation of ${r_{p,r,u}} $ can be calculated as:
\begin{eqnarray}\label{e7-16}
E(r_{p,r,u})_{HB} &=& E\left (\sum\limits_{i = 0}^{K - 1} {t_m^i} t_n^i\right )=E(t_m^0t_n^0) + E(t_m^1t_n^1) + ...E(t_m^{K - 1}t_n^{K - 1})\nonumber\\
&=&a  \cdot b  \cdot P_r(t_m^0 =a;t_n^0 = b)+...+a  \cdot b  \cdot P_r(t_m^{K-1} =a;t_n^{K-1} = b).
\end{eqnarray}
For the pixels at the same location of the blue noise ${\rm CA}_{\hexagon}$ have the same probability distribution when taking $K$ snapshots, we have $P_r(t_m^0 =a;t_n^0 = b)=...=P_r(t_m^{K-1} =a;t_n^{K-1} = b)$. Then,
\begin{eqnarray}\label{e7-17}
E(r_{p,r,u})_{HB} 
&=& K\cdot a  \cdot b  \cdot P_r(t_m^i = a;t_n^i = b)
=K\cdot a  \cdot b  \cdot P_r(t_m^i = a|t_n^i = b ) \cdot P_r(t_n^i = b)\nonumber\\
&=&K\cdot\left [a  \cdot b  \cdot P_r( t_m^i = a|t_n^i = b, d > \lambda _b) \cdot P_r(d > \lambda _b)\right.\nonumber\\
&&+\left.a  \cdot b  \cdot P_r(t_m^i = a|t_n^i = b, d < \lambda _b) \cdot P_r(d < \lambda _b)\right ]
\cdot P_r(t_n^i = b),
\end{eqnarray}
where $0 < a,b < 1$. When $t_n^i = b \ne 0 $, that means one of the three adjacent elements in the macro pixel $A$ has value 1. We denote this one-valued element as $Q$. 

If $d<\lambda _b$, for simplicity, we suppose the distances between $Q$ and the three hexagonal elements in $B$ are all smaller than $\lambda _b$. Then, the elements in $B$ should be all zeros and $t_m^i=0$. Since $0 < a < 1$, we have
\begin{eqnarray}\label{e7-18.5}
P_r(t_m^i = a| t_n^i = b,d < {\lambda _b}) = 0.
\end{eqnarray}
If $d>\lambda _b$, for simplicity, we suppose the distances between $Q$ and the three hexagonal elements in $B$ are all larger than $\lambda _b$. Then, the probability that $B$ includes one non-zero element is $g$ when $0 < g \le {1 \over 2}$, and the probability becomes $1-g$ when ${1 \over 2} < g \le 1$.
In general, 
\begin{eqnarray}\label{e7-18}
P_r(t_m^i = a| t_n^i = b,d > {\lambda _b})  =
\left\{ \begin{array}{l}
g,\;\textrm{for} \; 0 < g \le \frac{1}{2}\\
1 - g,\;\textrm{for} \; \frac{1}{2} < g \le 1.
\end{array} \right.
\end{eqnarray}
Substituting Eqs.~(\ref{e7-18.5}) and (\ref{e7-18}) to Eq.~(\ref{e7-17}), we have, 
\begin{eqnarray}\label{e7-19}
E(r_{p,r,u})_{HB}=K\cdot a \cdot  b\cdot  P_r(t_m^i= a|t_n^i = b,d > {\lambda _b} )\cdot  P_r(d > {\lambda _b}) \cdot  P_r(t_n^i= b).
\end{eqnarray}
Suppose $P_r(t_n^i= b)=g$, where $g$ is the transmittance of the coded aperture.
When $0 < g \le \frac{1}{2}$,
\begin{eqnarray}\label{e7-21}
E(r_{p,r,u})_{HB} =K\cdot a \cdot  b\cdot {g^2} \cdot  {P_r}(d > {\lambda _b}) <K\cdot{g^2} \cdot  {P_r}(d > {\lambda _b})=E(r_{p,r,u})_{SB}.
\end{eqnarray}
When $\frac{1}{2} < g \le 1$,
\begin{eqnarray}\label{e7-23}
E(r_{p,r,u})_{HB} &=& K\cdot a \cdot  b\cdot  g(1 - g) \cdot  {P_r}(d > {\lambda _b}) \nonumber\\
&<&K\cdot a \cdot  b\cdot{g^2}\cdot{P_r}(d >{\lambda _b})<K\cdot {g^2} \cdot  {P_r}(d> {\lambda _b})=E(r_{p,r,u})_{SB}.
\end{eqnarray}
In summary, we have
\begin{eqnarray}\label{e7-24}
{E(r_{p,r,u})_{SR}>E(r_{p,r,u})_{SB}}.
\end{eqnarray}

\section*{Funding}
Fundamental Research Funds for the Central Universities (2018CX01025); China Scholarship Council (201706035012); the US National Science Foundation and Intel Corporation under the Visual and Experiential Computing Initiative (1538950).

\end{document}